\documentclass[aps,prl,a4paper,superscriptaddress,twocolumn,showpacs,floatfix]
{revtex4} 
\usepackage{graphicx} 
\usepackage{graphicx,epsfig} 
\begin{document} 
\title{Strongly Correlated Superconductivity and Pseudogap Phase 
near a multi-band Mott Insulator}

\author{Massimo Capone}
\affiliation{Enrico Fermi Center, Rome, Italy}
\affiliation{Universit\`a di Roma ``La Sapienza'' and Istituto 
Nazionale per la Fisica della Materia (INFM), SMC and UDR Roma 1, 
Piazzale Aldo Moro 2, I-00185 Roma, Italy} 
\author{Michele Fabrizio} 
\affiliation{International School for Advanced Studies (SISSA), and Istituto 
Nazionale per la Fisica della Materia (INFM) UDR-Trieste SISSA, Via Beirut 2-4,
I-34014 Trieste, Italy} 
\affiliation{The Abdus Salam International Centre for Theoretical Physics 
(ICTP), P.O.Box 586, I-34014 Trieste, Italy} 
\author{Claudio Castellani}
\affiliation{Universit\`a di Roma ``La Sapienza'' and Istituto 
Nazionale per la Fisica della Materia (INFM), SMC and UDR Roma 1, 
Piazzale Aldo Moro 2, I-00185 Roma, Italy} 
\author{Erio Tosatti}
\affiliation{International School for Advanced Studies (SISSA), and INFM 
Democritos National Simulation Center, Via Beirut 2-4, I-34014 Trieste, Italy} 
\affiliation{The Abdus Salam International Centre for Theoretical Physics 
(ICTP), P.O.Box 586, I-34014 Trieste, Italy} 
\affiliation{Laboratoire de Mineralogie-Cristallographie de Paris,
Universite Pierre et Marie Curie, 4 Place Jussieu, 75252 Paris, France}

\date{\today} 
\begin{abstract} 

Near a Mott transition, strong electron correlations
may enhance Cooper pairing. This is demonstrated
in the Dynamical Mean Field Theory solution of a 
twofold-orbital degenerate Hubbard model with an inverted on-site Hund rule
exchange, 
favoring local spin-singlet configurations. 
Close to the Mott insulator, which here is a local version of a valence bond 
insulator, a pseudogap non-Fermi-liquid metal,
a superconductor, and a normal metal appear, in striking similarity with 
the physics of cuprates. The strongly correlated 
s-wave
superconducting state
has a larger Drude weight than the corresponding normal state.
The role of the impurity Kondo problem is underscored.

\end{abstract} 

\maketitle 
How superconductivity emerges so spectacularly out of a weakly doped Mott 
insulator is one of the fascinating and still controversial aspects of high 
T$_c$ cuprate superconductors. A novel {\sl strongly correlated
superconductivity} (SCS) scenario has been recently proposed 
\cite{Capone-Science} which deals with an ultimately related basic issue, 
namely the conditions
under which  Cooper pairing can be {\em enhanced}, 
rather than depressed, 
by strong electron repulsion. 
The key of the SCS proposal is the presence of 
pairing interaction term $J$, which is weak
but is not suppressed by the strong short-range repulsion $U$.
This is realized for an attraction which involves mainly spin and orbital 
degrees of freedom, which are not frozen near a Mott metal-insulator 
transition (MIT). In addition, close to the MIT, correlations slow down 
electron motion so much that the effective quasiparticle bandwidth becomes 
extremely small $W_{qp}\ll W$, $W$ being its uncorrelated value.
In these conditions the pairing attraction
can eventually equal the quasiparticle bandwidth $J\sim W_{qp}$. That 
drives the system to an intermediate-strong coupling superconducting regime 
where the {\sl maximum} superconducting gap $\Delta \sim J$ for given 
value of $J$ is reached, as opposed to the much smaller uncorrelated BCS 
value $\Delta_{BCS} \sim W \exp(-W/J)$.
A first theoretical realization of SCS was demonstrated in 
Ref.~\cite{CaponePRL,Capone-Science} by a Dynamical Mean Field Theory 
(DMFT)\cite{DMFT} solution of a three-fold degenerate 
model for tetravalent fullerides. 
It was shown that a narrow SCS region arises next  
the MIT, and is indeed characterized by a superconducting gap three 
orders of magnitude larger that the value one would obtain for the
same attraction in the absence of correlation.

DMFT maps a lattice model onto an Anderson impurity (AI) model 
subject to a self-consistency condition. 
Within this mapping SCS emerges\cite{CaponePRL}
through a competition between Kondo screening of the impurity (leading to a 
normal Fermi liquid) and an intra-impurity mechanism which 
forms a 
local non-degenerate singlet (a kind of local 
resonating valence bond state\cite{PWA}).  
This requires orbital degeneracy and inversion 
of Hund's rules, both ingredients present in the model of 
Ref.~\cite{Capone-Science}.
More recently a simpler AI with 
only 
twofold orbital degeneracy 
was shown by Wilson Numerical 
Renormalization Group (NRG) to display anomalous properties\cite{DeLeo}, 
suggesting its lattice generalization as a new 
candidate for a SCS.

\begin{figure}[t] 
\vspace{-1.5cm}
\centerline{ 
\includegraphics[width=8cm]{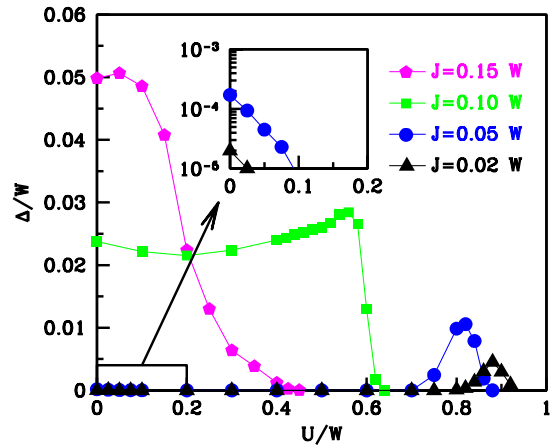} 
} 
\vspace{-1cm}
\caption{Superconducting gap for $\langle n \rangle =2$ as a function of
Hubbard repulsion for fixed coupling $J$=  $\lambda/ (2 N_F)$ . Increasing 
repulsion spoils superconductivity at large coupling. 
At weak coupling superconductivity is instead strongly enhanced 
close to the Mott transition.
The inset shows the weak coupling regime on an expanded scale, 
showing a much smaller gap at small $U$ compared with SCS at large $U$. 
\label{gap_n2}
} 
\end{figure}

In this Letter we present a detailed DMFT analysis that confirms this 
expectation, exploiting the lower degeneracy 
%
for
a wider and more revealing study. We consider an infinite coordination Bethe lattice and 
solve the AI by 
exact diagonalization. The model reads \cite{DeLeo}
\begin{equation}
\hat{H} = -t\sum_{<ij>,a,\sigma} 
c^\dagger_{i,a\sigma} c^{\phantom{\dagger}}_{j,a\sigma} + H.c. + \frac{U}{2}\sum_i n_i^2 + \hat{H}_J,
\label{Ham}
\end{equation}
where $c^\dagger_{i,a\sigma}$ creates an electron at site $i$ in orbital 
$a=1,2$ with spin $\sigma$, while $n_i= \sum_{a,\sigma} c^\dagger_{i,a\sigma} 
c^{\phantom{\dagger}}_{i,a\sigma}$ is the on-site occupation number. 
The on-site exchange is 
\begin{equation}
\hat{H}_J = -2J\, \sum_i \left( T_{i,x}^2 + T_{i,y}^2\right),
\label{HK}
\end{equation}
where $T_{i,\alpha} = 1/2 
\sum_{a,b}\sum_\sigma c^\dagger_{i,a\sigma}\, 
\left(\hat{\tau}^\alpha\right)_{ab}\, 
c^{\phantom{\dagger}}_{i,b\sigma}$ are the pseudo-spin 1/2 operators,
and
$\hat{\tau}^\alpha$ ($\alpha=x,y,z$) the 
Pauli matrices. The electronic states of the isolated site with $n$
electrons are labeled by total spin and pseudo-spin, $S$ and $T$
and their $z$-components, and have 
energies $E(n,S,S_z,T, T_z) = U n^2/2 - 2J [ T(T+1) - T_z^2]$. 
For $n=2$ the configurations allowed by Pauli principle 
are a spin triplet orbital singlet ($S=1$ and $T=0$) and a spin singlet 
orbital triplet ($S=0$ and $T=1$), split by $\hat{H}_J$. 
If $J<0$, standard Hund's rules, the spin triplet has the lowest energy.
Here we consider the less common case of $J>0$, where the lowest 
energy configuration has $S=0$, $T=1$ and $T_z=0$. This inversion of Hund's 
rules may for instance mimic a dynamical $e\otimes E$ Jahn-Teller 
effect\cite{Capone-Science,CaponePRL}. Here it 
just represents a generic mechanism for on-site 
spin-singlet pairing. Indeed, for $U=0$ and $J\ll W$, 
the ground state of (\ref{Ham}) is an $s$-wave BCS superconductor with pairs 
condensed in the $S=0$, $T=1$ and $T_z=0$ channel. The energy gap is
$\Delta\sim W \exp(-1/\lambda)$, where $\lambda = 2 J N_F$ is the 
dimensionless superconducting coupling, and $N_F$ the density of states (DOS) 
at the Fermi energy per spin and orbital. In this regime, 
a finite $U\ll W$ introduces a ``Coulomb'' pseudopotential $\mu_* = U N_F$ 
that opposes 
superconductivity, eventually suppressed for $U>2J$ in favor 
of a normal metal ground state. For 
larger $U\geq U_c\sim W$, the 
model (\ref{Ham}) undergoes a MIT for all integer fillings $\langle 
n\rangle=1,2,3$. We could generally expect the superconducting gap to 
decrease monotonically as a function of $U$, 
the superconductor either turning directly into a Mott insulator 
(for $\lambda\sim 1$), or (when $\lambda \ll 1$) first into a metal, 
and then into the insulator. This is indeed 
what we find for              
$\langle n \rangle =1$ (equivalent to $\langle n\rangle= 3$ by particle-hole 
symmetry) and $\lambda$ ranging from 0.1 to 0.6.  
The picture is however richer in the half-filled  
$\langle n \rangle =2$ case, as reported in Fig. ~\ref{gap_n2}, where
the standard superconductor evolves continuously to the SCS regime 
If $\lambda$ is large (e.g.,$J/W=0.15$), $\Delta$ decreases with increasing 
$U$, as in a BCS superconductor, until a weakly {\sl first order} transition 
to our local-RVB insulator occurs\cite{Gunnarsson}. 
For smaller $\lambda$ however, the dependence of $\Delta$ on $U$ is 
non-monotonic, and the initial drop for small $U$ is followed by a rise. 
Finally, for the smallest values of $\lambda$, the weak-$U$ superconductor
first turns into a metal but, just before the MIT, reverts back to a
superconductor, as in SCS of Ref.~\cite{Capone-Science}. Here too 
the superconducting gap 
reaches values much larger than those attained for the same pairing 
attraction $\lambda$ but 
for $U$ = 0 
(compare the inset and the right-hand side of  
Fig.~\ref{gap_n2}).

\begin{figure}[t]
\vspace{-2cm} 
\centerline{ 
\includegraphics[width=8cm]{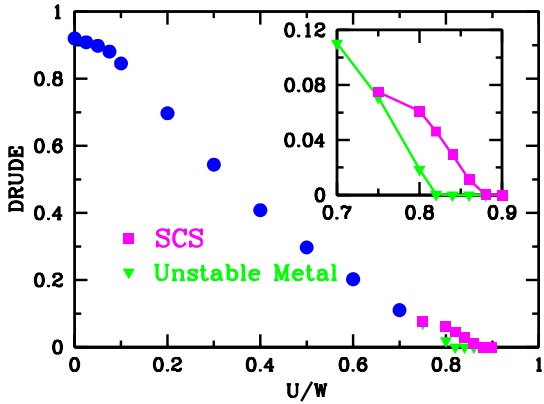} 
} 
\vspace{-0.5cm}
\caption{Drude weight as function of $U$ at $J=0.05 W$. SCS points:
lowest energy superconducting solution; unstable metal points : metastable
solution with superconducting order parameter forced equal to zero. Note
the very large Drude weight increase in the SCS pocket (magnified in the inset).
\label{Drude}}
\end{figure}

The appearance of SCS in (\ref{Ham}) for $\langle n\rangle =2$ 
supports the AI analysis~\cite{DeLeo}. Indeed, 
as $U$ increases, the quasiparticle bandwidth, which is the AI effective Kondo 
temperature $T_K$, 
is gradually suppressed by the 
dropping
quasiparticle 
residue $Z \ll 1$, $T_K\sim Z W$. 
%
At the same time,
we expect $J$ to remain 
unrenormalized~\cite{Capone-Science} . $H_J$ splits 
multiplets with same $n$, 
and that can 
happen
without opposing $U$, which just 
freezes charge 
fluctuations 
regardless of
either spin or orbital degrees of freedom. 
As $Z$ decreases, the AI enters the critical region around 
its unstable fixed-point \cite{DeLeo} at a critical 
$T_K^{(c)}\sim J$.  For $T_K>T_K^{(c)}$, the AI is in the Kondo screened 
regime, while for $T_K<T_K^{(c)}$ 
an intra-impurity singlet forms thanks to the inverted exchange $J$.
At the fixed point the two effects balance exactly,
leading to a residual entropy $\ln
\sqrt{2}$.
The remaining impurity degrees of freedom are quenched at a larger temperature 
$T_{+}\sim max(T_K,J)$.  $T_-\sim |T_K-T_K^{(c)}|^2/T_K^{(c)}$
measures the deviation from the fixed point behavior. 
It was argued in Ref.~\cite{DeLeo} that DMFT self-consistency should 
turn this AI instability into a true bulk one, most likely 
implying superconductivity. This is now confirmed by our DMFT analysis. 
S-wave superconductivity opens up a new screening route which freezes the 
residual entropy, otherwise quenched only below $T_-$\cite{DeLeo}. 
This suggests that (a) the energy scale which controls superconductivity 
should 
be related to $T_{+}-T_{-}$; (b) the onset of superconductivity 
should be
accompanied by a kinetic energy gain at low frequency, as Kondo 
screening  implies.  A kinetic energy (or, more accurately, 
band energy) gain may be regarded as a signal of SCS (as opposed to BCS
where kinetic energy rises), 
and that is reflected by the behavior of the Drude weight. 
In Fig.~\ref{Drude} we show for $J= 0.05~W$ the Drude weight of 
the stable superconducting solution for $0.7 \leq U/W\leq U_c$ (here the Drude 
weight  is the strength of the superfluid peak), in comparison with that of 
the {\sl unstable} metallic solution (obtained disallowing superconductivity),
taken to represent the normal phase.
Fig.~\ref{Drude} shows a large increase of Drude weight with 
superconductivity, an occurrence also predicted\cite{Hirsch, Baskaran} 
and actually observed in 
cuprates\cite{VanDerMarel}.
Here the unstable metal Drude weight actually appears 
to vanish at $U_* < U_c$.  As we shall show, this reflects the opening of a 
pseudogap at the chemical potential before the MIT. 

\begin{figure}[t] 
\centerline{ 
\includegraphics[width=6cm]{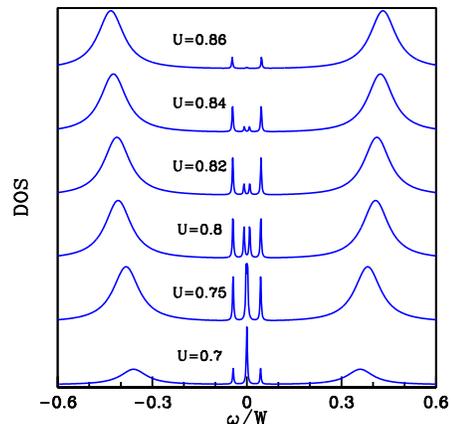} 
} 
\vspace{-0.5cm}
\caption{Single-particle DOS at $J=0.05 W$ of the metallic solution for 
different $U$'s close to the MIT (superconductivity between $U= 0.7$ and 
$U_c\simeq 0.88$ is not allowed).  Note the pseudogap within the low-energy 
quasiparticle peak coexisting with the high-energy Hubbard bands.
\label{DOS6}
} 
\end{figure}

\begin{figure}[b]
\vspace{-2cm} 
\centerline{ 
\includegraphics[width=8cm]{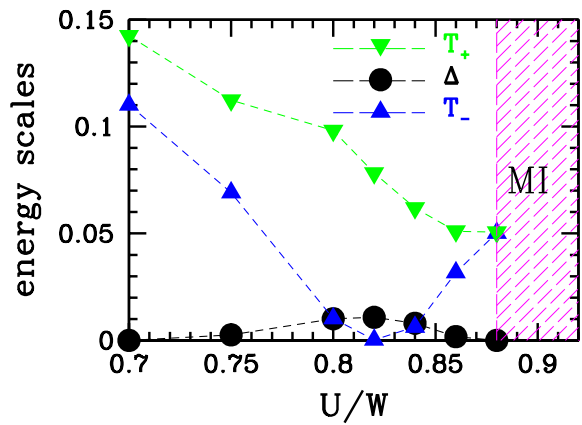} 
} 
\vspace{-0.5cm}
\caption{Behavior of the relevant energy scales (defined in the text)
close to the MIT for $J/W=0.05$ as function of $U/W$. 
\label{scale}
} 
\end{figure}

Additional physical insight is gained by analyzing the DOS 
in the strongly correlated region $U/W > 0.7$. Within DMFT, the DOS in this
region of parameters displays three features: Two high-energy Hubbard bands and
a low-energy feature associated to quasiparticles.
Fig.~\ref{DOS6} displays the evolution of the DOS of the normal state solution
in this regime, and shows that a pseudogap opens within the low-energy peak.
A similar pseudogap appeared also in the NRG solution of the AI
of Ref. \cite{DeLeoDOS}, where it was argued that 
the low-energy DOS around the unstable fixed 
point is well described by 
\begin{equation}
\rho(\omega) = \frac{\rho_0}{2}\left( \frac{T_+^2}{\omega^2+T_+^2} 
\pm \frac{T_-^2}{\omega^2+T_-^2}\right).
\label{AI:DOS}
\end{equation} 
where $\rho_0$ is the non-interacting DOS, 
and plus/minus refers to the Kondo screened/unscreened phase.
We find that  (\ref{AI:DOS}) fits well also the low-frequency behavior of 
the DMFT results\cite{nota-semicircolare}.
Eq.~(\ref{AI:DOS}) implies that at the chemical potential $\rho(0)=\rho_0$ 
in the Kondo screened regime, compatible with a Fermi-liquid with
${\cal I}m \Sigma(\omega)\sim \omega^2$.
On the contrary, the DOS values at $T_-=0$, $\rho(0)=\rho_0/2$, and
in the pseudogap 
phase, $\rho(0)=0$, imply a singular 
behavior of the self-energy, {\sl i.e.}, a non-Fermi liquid. 
The best-fit values of $T^+$ and $T^-$ for $J=0.05~W$ are shown in 
Fig.~\ref{scale}, and compared with the superconducting gap $\Delta$.
The maximum of $\Delta$ almost coincides with the vanishing of $T_-$,
which corresponds to the AI unstable fixed point.  
Fig.~\ref{scale} suggests a scenario which shares  
similarities to the quantum critical point\cite{quantum-critical-point} 
and the gauge theory-slave boson\cite{slave-bosons} pictures of
cuprates. For temperatures $T$ above $T_c$ (presumably of order $\Delta$), 
but below $T_-$, the normal phase is a Fermi-liquid
for $U < U_*$, and a non Fermi-liquid pseudogap phase for $U_c > U > U_*$.
For $T_- < T < T_+$, both the narrow Fermi-liquid quasiparticle 
peak and the pseudogap should be washed out, leaving a broader resonance
reflecting the properties of the non Fermi-liquid AI
unstable fixed point. The resonance will eventually disappear above 
$T_+$. Close to the MIT, the unstable metallic solution and the stable 
superconductor have almost the same energy and very similar DOS, but for the presence of 
a very small superconducting gap. 
From this point of view, the pseudogapped unstable phase plays 
a role similar to the staggered flux phase within the SU(2) invariant
slave-boson description of the $t$-$J$ model, and may  therefore be thought
as a phase with broken symmetry in the particle-hole or in
the particle-particle channels, where the full symmetry is restored by 
fluctuations\cite{slave-bosons}.
We note here that (a) 
superconductivity is not an accidental route which the lattice system takes 
to rid itself of competition among other phases, but is one of the  
pre-determined instabilities of the AI; (b) attempts to uncover the fixed 
point by suppressing superconductivity would likely spoil the fixed point or 
unveil other instabilities of the lattice model. 

The analogy with high T$_c$ cuprates becomes even more suggestive
when we analyze the phase diagram away from half-filling, 
and follow the fate of superconductivity upon doping. 
\begin{figure}[t]
\centerline{ 
\includegraphics[width=8cm]{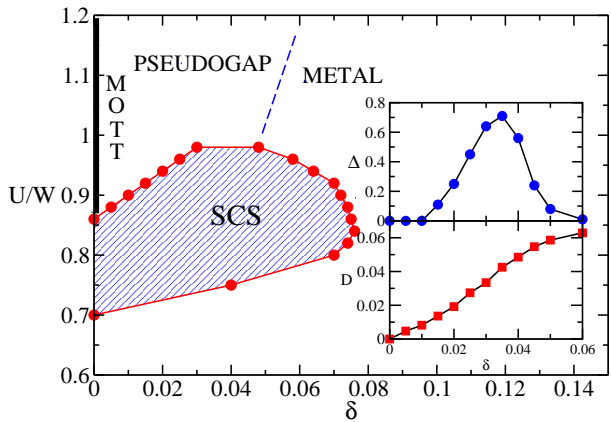} 
} 
\caption{Phase diagram as function of $U/W$ and doping $\delta=n-2$ at 
$J=0.05~W$. The think vertical line marks the 
singlet
Mott insulator. The inset shows, for $U=0.92W$, the superconducting gap 
$\Delta$ divided by a factor $10^{-3}$ and Drude weight $D$ (normalized to the
 non-interacting value) as functions of doping $\delta$.                  
\label{phd}} 
\end{figure} 
SCS extends (Fig.~\ref{phd}) into a superconducting pocket away from $\langle 
n\rangle =2$. When we dope the 
singlet 
Mott insulator just above the MIT, a pseudogap 
metallic phase is encountered first, followed by a narrow superconducting 
region which finally gives way to a normal metal (inset of Fig.~\ref{phd}), 
in remarkable similarity to the phase diagram of high T$_c$ cuprates. 
Striking is also the similarity in the Drude weight (inset), which is zero in 
the Mott insulator at half-filling and increases almost linearly upon doping.
The phase diagram of Fig.~\ref{phd} again agrees with the AI 
analysis\cite{DeLeoDOS}, which shows that the unstable 
fixed point still exists away from half-filling.

In the alkali fulleride models, that provide our prime examples of SCS, 
the physics which leads to $J >0$ and spin pairing is Jahn Teller, {\sl i.e.} 
a conventional phonon effect.\cite{CaponePRL,Capone-Science, Han}
Future experiments to test the SCS scenario by comparing 
Drude weights in presence and in absence of superconductivity should be 
actively considered in trivalent fullerides such as K$_3$NH$_3$C$_{60}$.\cite{NH3} 
These materials are of extraordinary importance, for they show a similar 
``universal'' contiguity between superconductivity and Mott insulating 
behavior as the high T$_c$ cuprates do, albeit with different 
physical ingredients.  When dealing with spectroscopies, one should 
keep in mind that there are now two low-energy scales, $T_+$ and $T_-$. 
That makes, e.g., photoemission data such as the recent 
ones on K$_3$C$_{60}$\cite{Goldoni} especially difficult to analyze, 
even ignoring the heavy vibronic structures. In the SCS region,
coherent band-like dispersion should occur with a reduced bandwidth
of order $2 T_+$, a scale which {\em remains finite} across the MIT,
even though the system is strongly correlated. The
true quasiparticle peak will instead become extremely narrow, of order
$T_-$, or even disappear in the pseudogap regime. 
We believe that the data\cite{Goldoni} are actually compatible with the SCS scenario.
The underlying Kondo physics offers a new
intriguing insight in this superconductivity problem. 
It seems in fact that the onset of bulk coherence 
close to a MIT may involve a hierarchy of energy scales, 
$T_- < T_+$ and $\Delta$. The latter is the 
most sought-after, as it marks the onset of lattice long-range order. 
On the contrary, $T_+$ and $T_-$ have their clearest meaning in the 
AI, which displays a two-stage quenching of the impurity entropy. 

We are grateful to L. De Leo and to J.E. Han for sharing with us their 
results prior to publication, and to P. Nozi\`eres, E. Cappelluti, 
G.E. Santoro and M. Lueders for illuminating discussions.
This work has been supported by MIUR Cofin 2003,
 MIUR FIRB RBAU017S8R004, FIRB RBAU01LX5H, and
by INFM Progetto Calcolo Parallelo.


\begin{thebibliography}{30}
\bibitem{nota-semicircolare} In our Bethe-lattice, 
the fit works better with semi-circular DOS's rather than lorentzians.
\end{thebibliography}
\end{document}